\documentclass[preprint,prd,amsmath,amssymb,aps,nofootinbib]{revtex4-1}

\usepackage{graphicx}
\usepackage{siunitx}
\usepackage{bm}
\usepackage{natbib}
\graphicspath{{figures/}}
\usepackage{threeparttable}

\begin{document}

\title{Resolving diffusion signatures in distant pulsar halos with current and future experiments}
\thanks{This work is supported by the National Natural Science Foundation of China under Grants No. 12393853, No. 12175248, and No. 12105292.}
\author{Yong-Jian Wei$^{a,b}$}
\email{weiyongjian@ihep.ac.cn}
\author{En-Sheng Chen$^{a,c}$}
\email{chenes@ihep.ac.cn}
\author{Kun Fang$^{a}$}
\email{fangkun@ihep.ac.cn}
\author{Xiao-Jun Bi$^{a,b,c}$}
\email{bixj@ihep.ac.cn}
\affiliation{
$^a$Key Laboratory of Particle Astrophysics, Institute of High Energy
Physics, Chinese Academy of Sciences, Beijing 100049, China \\
$^b$University of Chinese Academy of Sciences, Beijing 100049, China\\
$^c$TIANFU Cosmic Ray Research Center, Chengdu, Sichuan,  China
}


\date{\today}

\begin{abstract}
$\gamma$-ray pulsar halos are produced by electron-positron pairs that diffuse away from the pulsar and scatter off background photons. Their morphology serves as an ideal probe for studying cosmic-ray propagation on scales of several tens of parsecs. However, the number of firmly identified pulsar halos remains limited, primarily because current $\gamma$-ray experiments, limited by angular resolution, struggle to resolve the diffusion signatures of distant candidates ($>1$~kpc).
In this work, we investigate the prospects for identifying pulsar halo candidates through morphological discrimination using simulations of two advanced $\gamma$-ray experiments: the Kilometer Square Array of the Large High Altitude Air Shower Observatory (LHAASO-KM2A) and the Cherenkov Telescope Array (CTA, under construction). Using mock observations with realistic instrumental responses, we quantitatively assess the ability of each experiment to distinguish diffusion-based halo morphologies from alternative spatial models. 
Our analysis indicates that f the angular resolution of LHAASO-KM2A could be improved by $40\%$, it would be capable of resolving several prominent pulsar‑halo candidates, namely the halos around pulsars J1831‑0952, J0248+6021, and J0359+5414. CTA holds an advantage in resolving the morphology of sources beyond $\approx1.5$~kpc owing to its superb angular resolution. By extending exposure times to hundreds of hours, CTA is expected to achieve morphological identification for all known pulsar‑halo candidates.
\end{abstract}

\maketitle

\section{Introduction}
In 2007, the water Cherenkov experiment Milagro discovered $\gamma$-ray emission extending $\approx3^\circ$ around the bright $\gamma$-ray pulsar Geminga \cite{Abdo:2007ad}. This large extended halo was difficult to interpret by the pulsar itself or its associated bow-shock pulsar wind nebula \cite{2003Sci...301.1345C,Posselt:2016lot}. A decade later, the High-Altitude Water Cherenkov Observatory (HAWC) observed this halo with improved angular resolution and sensitivity, revealing that a diffusion model of electrons could well explain its spatial distribution \cite{HAWC:2017kbo}. This finding pointed to the most likely origin of the Geminga halo: electron-positron pairs ($e^\pm$) born in the pulsar escape into the interstellar medium (ISM) and, as they diffuse in the turbulent magnetic field, produce extended $\gamma$-ray emission through inverse Compton scattering with background photons. Consequently, a new class of astrophysical sources emerged, named pulsar halos or TeV halos. The morphology of a pulsar halo is a clear projection of the spatial distribution of its parent $e^\pm$, making it a unique probe for studying the cosmic-ray propagation in a specific ISM region \cite{Fang:2022fof,Liu:2022hqf,Lopez-Coto:2022igd,Amato:2024dss}.

Based on surveys of TeV $\gamma$-ray sources conducted by wide-field experiments HAWC and the Large High Altitude Air Shower Observatory (LHAASO), the number of confirmed or candidate pulsar halos has reached around ten \cite{Albert:2020fua,LHAASO:2023rpg}. However, apart from the Geminga halo and the Monogem halo \cite{HAWC:2017kbo}, the corresponding pulsars for other potential halos are all located at distances exceeding $1\ \text{kpc}$, resulting in significantly smaller angular extensions of these halos. In the case of LHAASO J0621$+$3755, the morphology measurements, affected by instrumental angular resolution, cannot distinguish between a diffusion model and mathematical templates such as disk or Gaussian profiles \cite{Aharonian:2021jtz}. The lack of clear diffusion signatures in the observed morphology complicates the confirmation of pulsar halos.

To reliably expand the sample size of pulsar halos, enhanced instrumental angular resolution plays a crucial role. The Cherenkov Telescope Array (CTA, \cite{CTAConsortium:2010umy}), currently under construction, will achieve unprecedented angular resolution ($<0.05^\circ$ above $10\ \text{TeV}$ \cite{Schwefer:2024bsx}) in $\gamma$-ray astronomy, along with significantly improved sensitivity compared to current atmospheric Cherenkov experiments. This will serve as the primary means of identifying diffusion signatures in distant pulsar halos. On the other hand, LHAASO, with its wide field of view and long exposure time, remains indispensable for precise observations of relatively nearby pulsar halos. In the future, LHAASO plans to integrate deep learning techniques to improve directional reconstruction, thereby achieving better angular resolution for resolving the morphologies of more distant pulsar halos. Based on these prospects, this study examines the potential of experiments to decipher diffusion signatures in candidate pulsar halos and evaluates how enhancements in angular resolution or exposure time can boost the identification.

The paper is organized as follows. Section \ref{sec:model} presents a simplified template of pulsar halos that effectively captures their morphological characteristics. Section \ref{sec:experiments} introduces the $\gamma$-ray experiments and the methods for simulating pulsar halo signals based on experimental information. Section \ref{sec:statistics} presents the statistical approach used for morphological discrimination. Section \ref{sec:result} demonstrates the capability of experiments to conduct morphological tests on pulsar halos. Section \ref{sec:conclu} is the conclusion.

\section{Diffusion-based Pulsar Halo Model}
\label{sec:model}
Electron-positron pairs produced by pulsars can be accelerated to energies as high as $100\ \text{TeV}$ by relativistic shocks in PWNe, after which they continuously escape into the ISM. When a PWN evolves into the bow-shock stage, its size remains on the order of $\sim1\ \text{pc}$ for an extended period under the confinement of the ISM ram pressure \cite{Gaensler:2006ua}, which is much smaller than the characteristic diffusion scale of the escaped $e^\pm$. Consequently, the propagation of the parent $e^\pm$ in pulsar halos can be regarded as a diffusion–radiation-loss problem under the point-source assumption. Under this scenario, the $\gamma$-ray surface brightness of a pulsar halo can be approximately described by the following parameterization \cite{HAWC:2017kbo}:
\begin{equation}
 \begin{aligned}
 f(\theta,E_{\gamma}) = & f_{0}\left(\frac{E_{\gamma}}{30~\text{TeV}}\right)^{-\alpha}e^{-E_{\gamma}/E_{c}} \times \frac{1.22}{\pi^{3/2}\theta_d(\theta+0.06\theta_d)}e^{-(\theta/\theta_d)^{2}}, \\
 \end{aligned}
\label{eq:theta_d}
\end{equation}
where $f_{0}$ is the normalization at $30~\text{TeV}$, $\alpha$ is the spectral index, $E_c$ is the cutoff energy, $\theta$ is the angular distance from the pulsar, and $\theta_d$ is the characteristic angular extension of the halo, which can be dependent on $E_\gamma$. The $\gamma$-ray spectrum is assumed to be a power-law with an exponential cutoff (ECPL) form, effectively reflecting that the parent $e^\pm$ originate from a shock-acceleration process.

The spatial distribution term in Eq.~(\ref{eq:theta_d}) exhibits a pronounced centrally-peaked profile, which differs significantly from the uniform disk template commonly used to describe the profile of supernova remnants. Additionally, a Gaussian template is also generally used in $\gamma$-ray astronomy to characterize the morphology of extended sources with a centrally-peaked profile. Yet, it also remarkably differs from Eq.~(\ref{eq:theta_d}): within the range $0.06\theta_d < \theta < \theta_d$, Eq.~(\ref{eq:theta_d}) decreases sharply with increasing $\theta$, exhibiting a steeper decline than the Gaussian distribution. However, all these distinctions may become blurred due to the insufficient instrumental angular resolution or the limited photon statistics.

In this work, we aim to investigate the capability of $\gamma$-ray experiments to identify pulsar halos through morphological discrimination. We adopt the diffusion-based parameterization (hereafter referred to as diffusion model) presented by Eq.~(\ref{eq:theta_d}) to describe the morphology of pulsar halos, and consider the disk and Gaussian models as representative alternatives for non-halo origins.

The variables that primarily reflect differences among pulsar halos are the pulsar distance $d$ and its spin-down luminosity $\dot{E}$. Generally, morphology discrimination becomes more difficult for more distant sources or those with lower spin-down luminosity. At present, the $\gamma$-ray energy spectrum and morphology of the Geminga halo have the most precise measurements. For simplicity, we assume that all the halos share the same spectral shape as the Geminga halo. The $\gamma$-ray energy spectrum of the Geminga halo is obtained by fitting the HAWC data \cite{HAWC:2024scl}, which yields $\alpha=1.49$ and $E_c=22.7$~TeV, as the benchmark model to generate mock data for all halo candidates. 

Additionally, we assume that the $f_0$ and $\theta_d$ of a pulsar halo scale depending on the ratio of its pulsar distance to that of Geminga: $f_0=f_{0,G}(d/d_G)^{-2}$, This scaling implies an underlying assumption that all pulsar halo candidates share the same pair-conversion efficiency from spin-down power as the Geminga pulsar halo.
$\theta_d=\theta_{d,G}(d/d_G)^{-1}$, where $\theta_{d,G}$ is set to $5.5^\circ$ as measured by HAWC \cite{HAWC:2017kbo}. For simplicity, $\theta_d$ is assumed to be energy independent.

We are aware that adopting a unified spectral shape and an energy-independent extension for all pulsar halos is a simplified assumption. However, known pulsar halos exhibit similar spectral shapes, and existing energy-dependent morphological measurements indicate that the extension of pulsar halos in the TeV regime does not vary dramatically \cite{HAWC:2024scl,LHAASO:2024flo}. This suggests that our assumption does not conflict with current observations.

\section{Instrumental Setup and Monte Carlo data}
\label{sec:experiments}
Ground-based observations of high-energy $\gamma$ rays are performed indirectly by measuring secondary particles generated in extensive air showers initiated by primary gamma rays. Two main experimental techniques are employed to detect such emission. The first is the Imaging Atmospheric Cherenkov Telescope (IACT) technique, which observes Cherenkov light flashes generated by relativistic shower particles propagating through the atmosphere during clear, moonless nights. The second is the extensive air shower (EAS) technique, which detects secondary shower particles directly at ground level. In this work, we focus on two representative instruments employing these approaches: CTA and the Kilometer Square Array of LHAASO (LHAASO-KM2A).

\subsection{CTA}
CTA will be the flagship next-generation instrument in $\gamma$-ray astronomy, covering photon energies from 20 GeV to 300 TeV. It is designed to achieve an energy resolution better than $10\%$ and a substantially improved angular resolution compared to existing $\gamma$-ray telescopes. CTA will operate as a global observatory with two sites: a Northern Hemisphere array (CTA-North) located on La Palma, Spain, and a Southern Hemisphere array (CTA-South) located at Paranal, Chile. In the so-called Alpha Configuration, CTA-North will consist of four Large-Sized Telescopes (LSTs) and nine Medium-Sized Telescopes (MSTs), while CTA-South will comprise 14 MSTs and 37 Small-Sized Telescopes (SSTs). The primary mirror diameters of the LSTs, MSTs, and SSTs are 23 m, 11.5 m, and 4.3 m, respectively. Their corresponding fields of view are larger than $4.5^{\circ}$, $7^{\circ}$, and $8^{\circ}$ \cite{CTAConsortium:2017dvg}.

Instrument Response Functions (IRFs)—including angular and energy resolutions, effective area, and expected background rates—are commonly used to derive differential sensitivities and to enable comparisons between different gamma-ray instruments. To facilitate a direct comparison with LHAASO-KM2A under similar sky coverage, we adopt the CTA-North IRFs for on-axis observations. These IRFs are publicly available through the Zenodo repository \cite{cherenkov_telescope_array_observatory_2021_5499840}. Following standard event selection procedures, including $\gamma$–hadron separation cuts, we obtain the photon effective area and residual background rates, consistent with the methodology described in Ref.~\cite{Celli:2024cny}.

\subsection{LHAASO-KM2A}
LHAASO is a new-generation $\gamma$-ray and cosmic-ray observatory located at an altitude of 4410 m in Daocheng, Sichuan province, China ($29^\circ 21' 31''$ N, $100^\circ 08' 15''$ E) \cite{LHAASO:whitepaper}. It consists of three sub-arrays: the LHAASO-KM2A, the Water Cherenkov Detector Array (LHAASO-WCDA), and the Wide Field-of-view Cherenkov Telescope Array (LHAASO-WFCTA). Partial operation of LHAASO began in 2019, and the full construction was completed in 2021. LHAASO-KM2A is composed of 5216 electromagnetic detectors (EDs) and 1188 muon detectors (MDs), covering an area of 1.3 $\rm{km}^2$. The EDs (MDs) are deployed with spacings of 15 m (30 m). The EDs are plastic scintillation detectors that measure the electromagnetic component of EASs and are designed primarily for event reconstruction. The MDs detect the muonic component of EASs and are mainly used for cosmic-ray background discrimination. High-energy $\gamma$-ray events above 10 TeV are predominantly recorded by KM2A. The array has a wide field of view of approximately 2 sr and operates with a nearly full duty cycle, providing unprecedented sensitivity for surveying the $\gamma$-ray sky at energies above 20 TeV.

We adopt the LHAASO-KM2A instrument response IRFs calibrated using observations of the Crab Nebula \cite{LHAASO:2024zug}. Specifically, we use the IRFs corresponding to a zenith angle of $\theta = 20^\circ$, ensuring consistency with those released by the other experiments considered in this work. The expected background rate is derived by convolving the all-particle cosmic-ray flux \cite{Gaisser:2013bla} with the effective area. The cosmic-ray (CR) effective area is derived from the reference $\gamma$-ray effective area by accounting for their respective selection efficiencies. Specifically, assuming the initial effective areas for CRs and $\gamma$-rays are identical prior to selection, the CR effective area is calculated as $A_{\text{eff}}^{\text{CR}} = (A_{\text{eff}}^{\gamma} / \epsilon_{\gamma}) \cdot \epsilon_{\text{CR}}$, where $\epsilon_{\gamma}$ and $\epsilon_{\text{CR}}$ denote the survival fractions for $\gamma$-rays and CRs after selection cuts, respectively \cite{LHAASO:2024zug}. This procedure ensures that the angular resolution, effective area, and expected background rate remain consistent with the benchmarks established in Ref.~\cite{Celli:2024cny}.

\subsection{Monte Carlo data}
\label{subsec:MC}
A direct comparison between the LHAASO-KM2A and CTA performance is now possible, since the same energy binning—namely 0.2 per decade in logarithmic scale—is adopted for both instruments. Based on their respective performance characteristics and the energy range relevant for halo spectral measurements, we consider energies from 10 TeV to 100 TeV for LHAASO-KM2A and from 1 TeV to 100 TeV for CTA.

Owing to its pointing observation mode and imaging camera, CTA achieves an excellent angular resolution. At 30 TeV, the $1\sigma$ width of the point spread function (PSF) is $\sigma_{\rm PSF} \sim 0.03^\circ$ for CTA, compared to $\sigma_{\rm PSF} \sim 0.3^\circ$ for LHAASO-KM2A. However, since CTA detects Cherenkov light produced by air showers, it can operate only under clear, dark-sky conditions. In contrast, LHAASO-KM2A has a nearly full duty cycle and can operate continuously, both day and night.

For a given extended halo, the differential flux at any spatial position and energy is determined by the assumed spectral and morphological models. The expected number of signal events, $N_s$, per energy bin and per spatial pixel within the region of interest (ROI) is calculated by folding the model flux with the instrument response functions, including PSF, and multiplying by the effective area, observation time, and solid angle. In this work, the model flux is defined by the diffusion morphology and the ECPL spectrum introduced in Section~2.1. The number of background events, $N_b$, is computed from the background rate, which is assumed to be constant in time. Using a toy Monte Carlo (toyMC) approach, the mock observed event counts, $N_{\rm on}$, are generated by drawing Poisson realizations of $N_s + N_b$. 

\section{Morphological Discrimination Method}
\label{sec:statistics}
For a given $d$ and $\dot{E}$, mock data can be generated based on the diffusion model through the toyMC technique outlined in Section~\ref{subsec:MC}. Next, the simulated data is fitted with the model to be tested using the maximum-likelihood method to determine the best-fit parameters. Subsequently, we evaluate the best-fit model using a specific statistical test to obtain a decision for this single trial.

We repeat the above test multiple times and compile the decision from each trial. The proportion of trials in which the diffusion model is correctly chosen (or the wrong model is rejected) relative to the total number of trials is defined as the power of the test. The power reflects the capability of the considered experiment to discriminate the diffusion model.

\subsection{Likelihood function}
In the ROI, a joint maximum-likelihood fit is performed over energy and spatial bins. Assuming Poisson statistics, the likelihood function is written as
\begin{eqnarray}
\ln \mathcal{L} = \sum_{i=1}^{m}\sum_{j=1}^{n}
\ln \left[
P\left(N_{ij}^{\rm on} \big| N_{ij}^{\rm b}+N_{ij}^{s}\right)
\right],
\label{eq:like}
\end{eqnarray}
where $P$ denotes the Poisson probability distribution, $i$ labels the $i$-th energy bin and $j$ the $j$-th spatial pixel. Here, $N_{ij}^{\rm on}$ is the number of observed events, while $N_{ij}^{\rm b}$ and $N_{ij}^{s}$ are the expected background and signal counts, respectively.

The signal expectation $N_{ij}^{s}$ is determined by the spatial model under test (Gaussian, disk, or diffusion) and the ECPL spectral model outlined in Section~\ref{sec:model}, folded with the instrument response functions. To simulate a realistic data analysis pipeline, we treat $f_0$, $\alpha$, $E_c$ and the angular width of the spatial model as unconstrained free parameters during the fitting process, identical to the procedure applied in the analysis of experimental data. The maximum likelihood is searched using the \texttt{MIGRAD} algorithm embedded in the Minuit Minimization package\footnote{\url{https://root.cern.ch/doc/master/group__MinuitOld.html}}. Larger $\ln \mathcal{L}$ values indicate a stronger preference for the presence of an extended source with the assumed morphology.

\subsection{Single trial}

We adopt two distinct statistical metrics for a single trial to discriminate among different spatial models: the Akaike Information Criterion (AIC) and the $\chi^2$ goodness-of-fit test. When data samples are sufficient, discriminating in favor of the diffusion model over a wrong model is generally easier than rejecting the wrong model. Therefore, the AIC typically yields higher statistical power than the $\chi^2$ test. However, compared to the AIC, the decision criterion of the $\chi^2$ test can be explicitly linked to a probability, which also provides valuable reference for our study.

\begin{figure}[!htb]
\begin{center}
\includegraphics[width=0.6\textwidth]{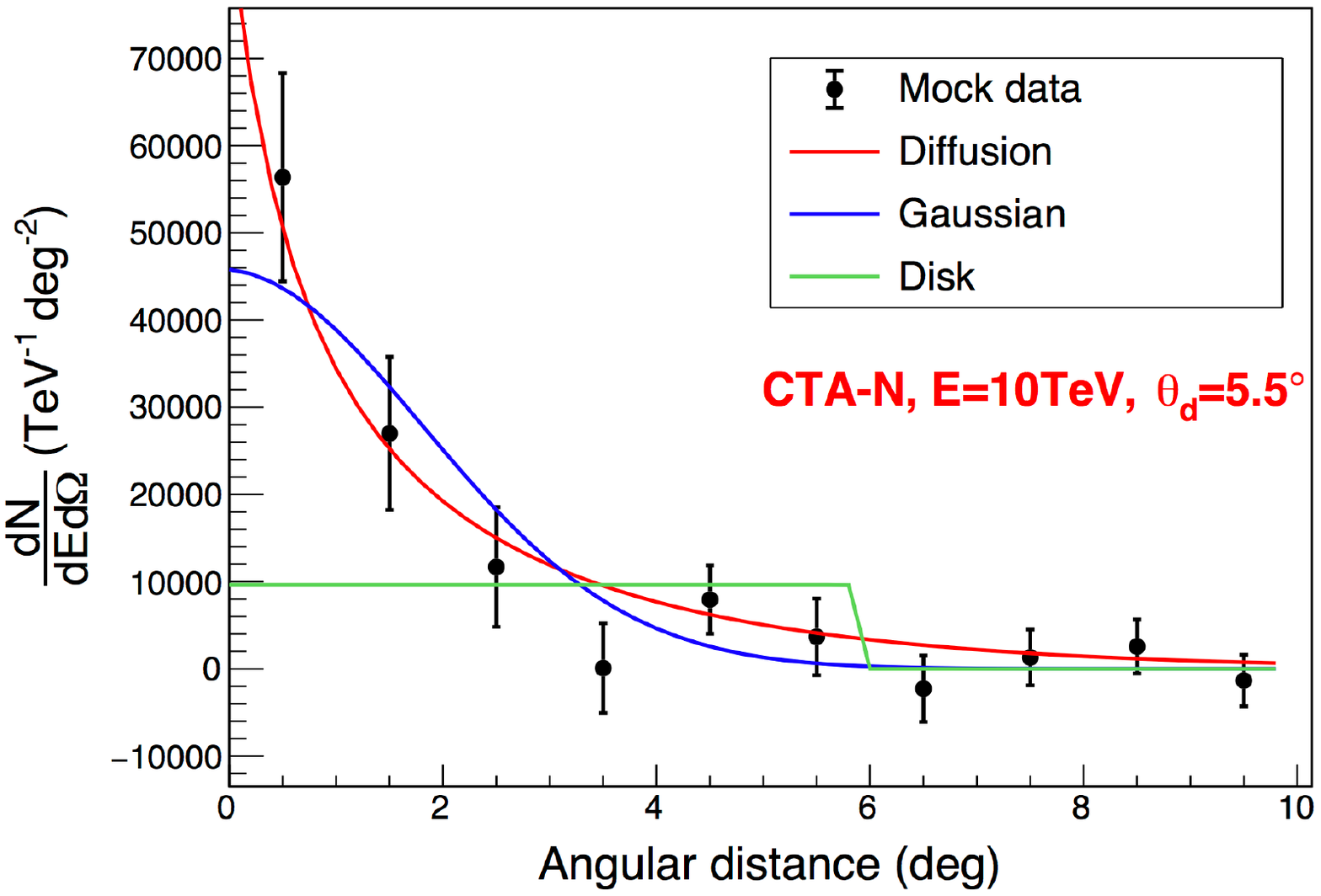}
	\caption{Radial $\gamma$-ray brightness profile of a pulsar halo with the Geminga parameters. The black points represent mock profile data of CTA-North ($50$~h) generated from the diffusion model. The red, blue, and green curves show the best-fit diffusion, Gaussian, and disk models, respectively.} 
    \label{fig:profile}
\end{center}
\end{figure}

\begin{enumerate}
 \item AIC: Since the diffusion model and the alternative models are non-nested, the likelihood ratio test and Wilks' theorem cannot be directly applied to assess the statistical significance of their differences. Therefore, we employ the AIC as a model selection metric, which is widely used in data analysis for comparing non-nested models. 
 The AIC is defined as
 \begin{equation}
    \mathrm{AIC} = 2k - 2\ln \mathcal{L}_{\max},
 \end{equation}
 where $k$ is the number of free parameters in the model and $\mathcal{L}_{\max}$ is the maximum likelihood. To compare different spatial hypotheses, we define
 \begin{equation}
  \Delta \mathrm{AIC} = \mathrm{AIC} - \mathrm{AIC}_{\rm diffusion},
  \label{eq:dAIC}
 \end{equation}
 where $\mathrm{AIC}_{\rm diffusion}$ corresponds to the diffusion model. Smaller values of $\Delta \mathrm{AIC}$ indicate a preference for the alternative model. We adopt $\Delta \mathrm{AIC} > 6$ as the criterion for model discrimination, which represents strong evidence \cite{burnham2002model, liddle2007information} in favor of the diffusion model.

 \item $\chi^2$ test: The $\chi^2$ statistic provides a straightforward and intuitive way to assess the compatibility between different spatial hypotheses and the mock data. Specifically, the $\chi^2$ test is performed on one-dimensional (1D) radial brightness profiles, with bin widths chosen to ensure at least 10 counts per bin. This threshold allows Poisson fluctuations to be reliably approximated by Gaussian statistics. For a given spatial template, the $\chi^2$ value is calculated as:$$\chi^2 = \sum_{i=1}^{n} \frac{(O_i - T_i)^2}{\sigma_i^2}$$where $O_i$ represents the observed excess in the $i$-th bin, $T_i$ is the corresponding model prediction, and $\sigma_i$ denotes the statistical uncertainty. When alternative spatial models, such as Gaussian or disk profiles, are fitted to the 1D radial brightness profiles generated by the diffusion model, they generally yield poorer $\chi^2$ values than the diffusion model, as shown in Fig.~\ref{fig:profile}. As the statistical quality of the data improves, these discrepancies become more pronounced, allowing incorrect models to be excluded based on their diminished goodness of fit. We adopt a $2\sigma$ exclusion criterion, corresponding to a p-value of approximately $0.05$ for the $\chi^2$ distribution with $n-k$ degrees of freedom, where $n$ is the number of spatial bins and $k$ is the number of free parameters in the fitted model. Models resulting in a $\chi^2$ value exceeding the critical threshold at this significance level are considered distinguishable from the diffusion scenario.

\end{enumerate}

\subsection{Statistical power threshold}
\label{subsec:power}
To mitigate the impact of statistical fluctuations arising from a single Monte Carlo realization, we perform 1000 independent toyMC simulations for each parameter configuration and obtain the corresponding statistical power. 

We establish a statistical power threshold of $0.8$, meaning that at least 800 out of 1000 realizations must satisfy the discrimination criterion (either the $\Delta \mathrm{AIC} > 6$ condition or the $2\sigma$ $\chi^2$ exclusion). When a specific combination of $d$ and $\dot{E}$ yields a statistical power of $0.8$, it indicates that the diffusion signature of the pulsar halo under these parameters can be resolved by the considered experiment. Consequently, we can determine a critical curve in the $d-\dot{E}$ plane, above which lies the parameter space where diffusion signatures can be distinguished.

\section{Results}
\label{sec:result}
\begin{figure}[!htb]
\begin{center}
\includegraphics[width=0.48\textwidth]{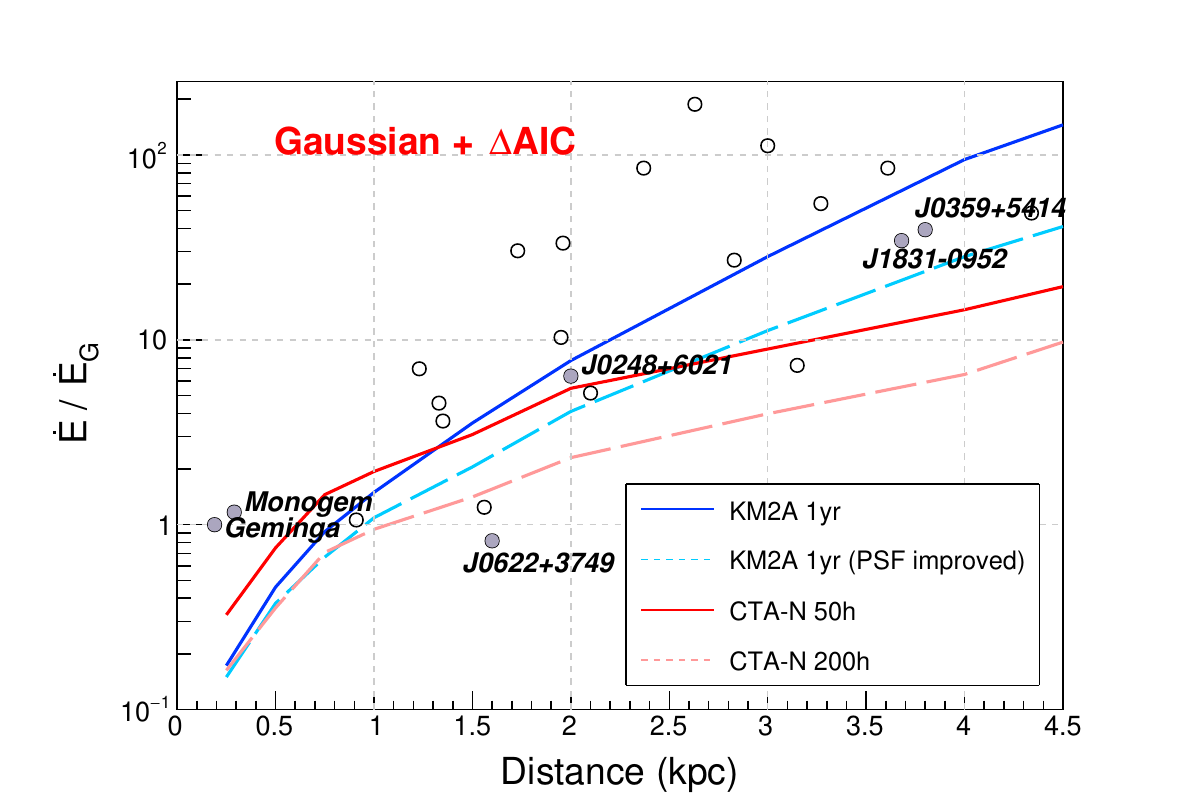}
\includegraphics[width=0.48\textwidth]{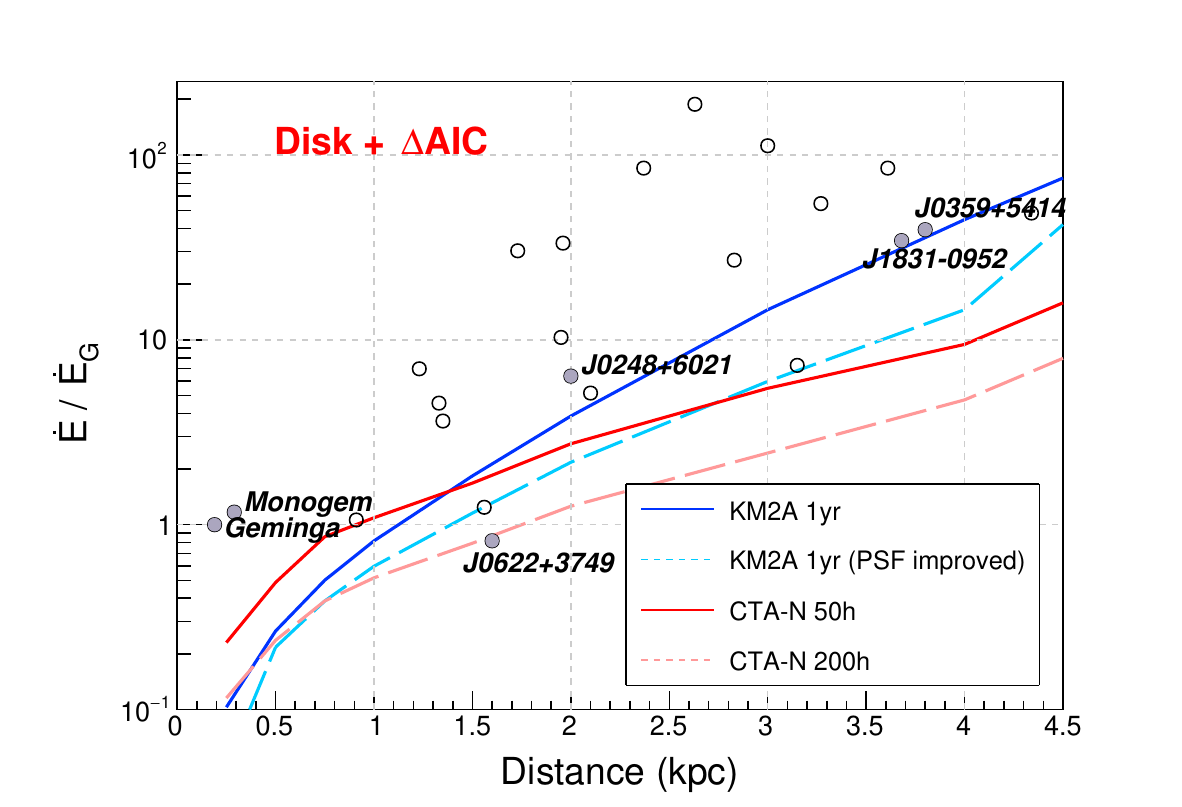}
\includegraphics[width=0.48\textwidth]{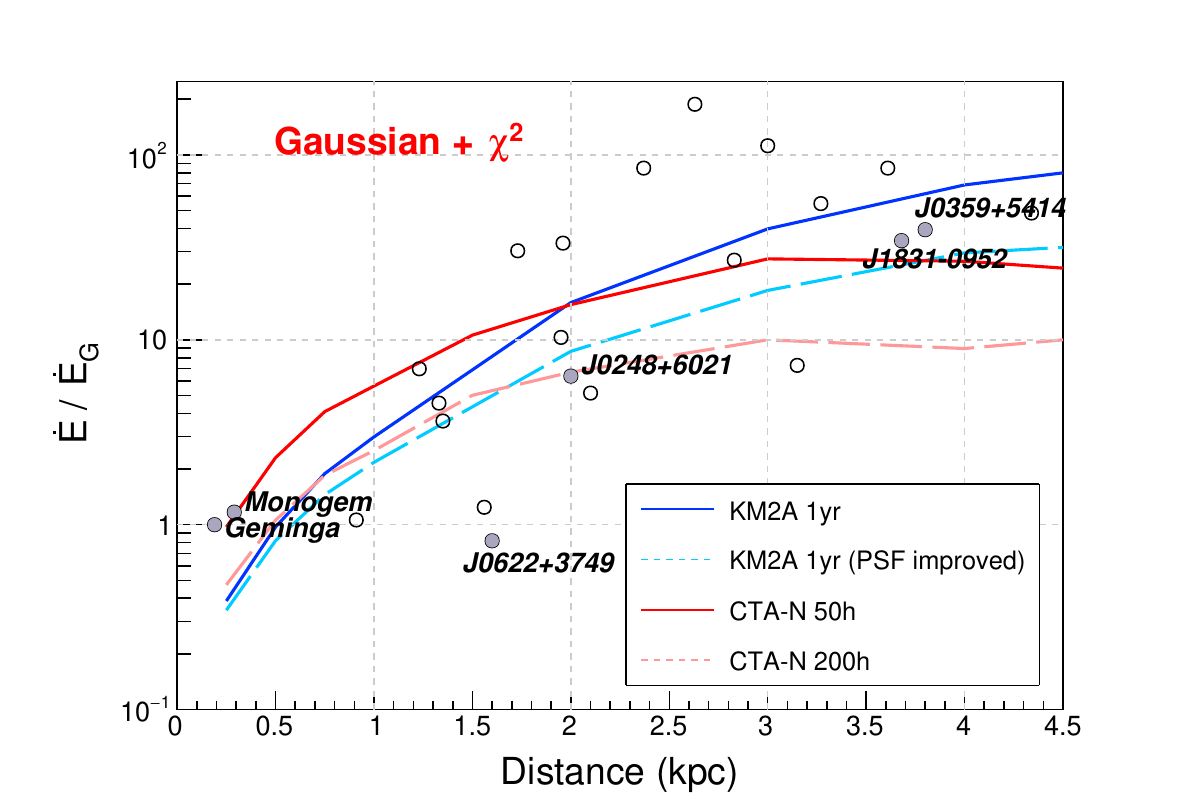}
\includegraphics[width=0.48\textwidth]{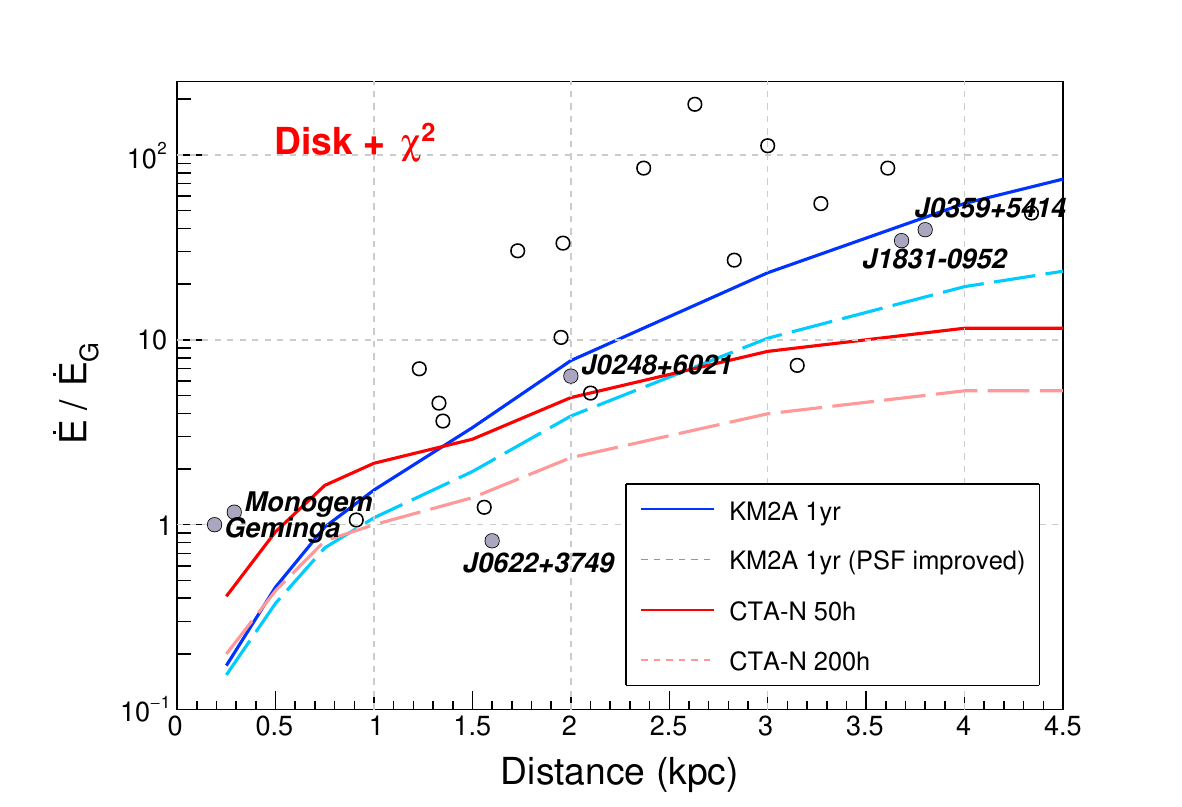}
	\caption{Discrimination curves for pulsar halos achievable with CTA-North and LHAASO-KM2A. We use $\dot{E}/\dot{E}_G$ instead of $\dot{E}$ as the vertical axis, where $\dot{E}_G$ denotes the luminosity of the reference source, Geminga. Sources located above the curves can be morphologically distinguished as diffusion halos. Filled circles denote confirmed pulsar halos or candidates that have already been reported in the literature, while empty ones denote other pulsar-associated sources in the LHAASO catalog. The competing spatial models and statistical decision methods used in the four panels, from top left to bottom right, are as follows: Gaussian model with AIC, disk model with AIC, Gaussian model with $\chi^2$ test, and disk model with $\chi^2$ test.} 
    \label{fig:discrimination_curve}
\end{center}
\end{figure}

The results of the morphological discrimination analysis are summarized in Fig.~\ref{fig:discrimination_curve}, i.e., the critical curves in the $d-\dot{E}$ plane introduced in Section~\ref{subsec:power}. The four subfigures correspond to different combinations of statistical criteria and competing spatial templates: the AIC and the $\chi^{2}$ test, each applied to comparisons between the diffusion-based halo model and Gaussian or disk spatial templates. In each subfigure, four discrimination curves are shown, representing different experimental configurations: LHAASO-KM2A (blue solid line), LHAASO-KM2A with PSF improved by $40\%$ (blue dashed line), CTA-North with 50~h exposure (red solid line), and CTA-North with 200~h exposure (red dashed line). Notably, for LHAASO-KM2A, the 1-year exposure time adopted in this study refers to the cumulative effective exposure time rather than the calendar timespan. Given that a source at $\delta = 30^\circ$ has an effective daily observation window of $7.8\,\mathrm{h}$ ($\theta_{\mathrm{zen}} < 50^\circ$), this 1-year effective exposure corresponds to a total observational period of approximately 3 calendar years. 

The positions of confirmed or potential pulsar halos are overlaid for reference, including the Geminga halo and the Monogem halo \cite{HAWC:2017kbo,HAWC:2024scl},  LHAASO J0621$+$3755 \cite{LHAASO:2021crt}, LHAASO J0249$+$6022 \cite{LHAASO:2024flo}, HESS J1831$-$098 \cite{Fang:2022qaf,2025arXiv251003183S}, and the source around PSR J0359$+$5414 \cite{HAWC:2023jsq}. The characteristic parameters of the pulsars associated with these candidates are summarized in Table \ref{tab:pulsar}. In addition, we consider pulsar-associated sources from the LHAASO catalog with characteristic ages greater than $20~\mathrm{kyr}$, excluding younger systems that may still be dominated by emission from initial pulsar wind nebulae. Halo candidates located outside the LHAASO field of view are not shown \cite{2025arXiv251002802W,2026arXiv260121689D}. It should be noted that the performance metrics for LHAASO-KM2A presented here are derived under optimal observation conditions, effectively representing the sensitivity for sources at $\delta \approx 30^\circ$. For sources at other declinations, the detection performance decreases, requiring longer exposure times to achieve comparable statistical significance. For instance, a source such as HESS J1831$-$098, located at a less favorable declination, would require at least twice the effective observation time to reach the same level of sensitivity.

\begin{table}[]
\centering
\begin{tabular}{cccccccc}
\hline 
Pulsar Name & (RA, DEC) & $P$ & $\Dot{P}$ & $\Dot{E}$ & Age & Dist  & Association\\ 
 & ($^\circ$,$^\circ$) & (s) & ($\rm{10^{-14}~s ~s^{-1}}$) & $\rm{(10^{34}erg ~s^{-1})}$ & (kyr) & (kpc) \\
\hline \text { J0633+1746 } & (98.48, 17.77) & 0.237 & 1.098 & 3.3 & 342.0 & 0.25  & Geminga \cite{HAWC:2017kbo,HAWC:2024scl}\\
\text { B0656+14 } &(104.95, 14.24)& 0.385 & 5.499 & 3.8 & 110.0 & 0.29 & Monogem \cite{HAWC:2017kbo,HAWC:2024scl}\\
\text {J0622+3749} &(95.54, 37.82)& 0.333 & 2.542 & 2.7 & 207.8 & 1.6 & LHAASO J0621+3755 \cite{LHAASO:2021crt}\\
\text { J0248+6021 } & (42.08, 60.36)&0.217 &  5.509 &  21.3  &  62.4  &  2.0   & LHAASO J0248+6021 \cite{LHAASO:2024flo} \\
 \text { J0359+5414 } &(59.86, 54.25)& 0.079 &  1.673  & 130  &   75   &   3.45 & HAWC J0359 \cite{HAWC:2023jsq}\\
 \text {J1831-0952} & (277.89, -9.87)&0.067 & 0.832 & 108 & 128 & 3.68 & HESS J1831-098 \cite{Fang:2022qaf,2025arXiv251003183S}\\
\hline
\end{tabular}
    \caption{Characteristic parameters of the pulsars associated with the halo candidates: RA and Dec represent Right Ascension and Declination in the equatorial coordinate system, $P$ is the spin period, $\dot{P}$ is the period derivative, $\dot{E}$ is the spin-down luminosity, Age denotes the characteristic age, and Dist represents the distance from Earth.}
    \label{tab:pulsar}
\end{table}

A general trend is that the AIC-based method yields less restrictive discrimination conditions than the $\chi^{2}$ test. For the same spatial template and instrumental configuration, the AIC criterion identifies a larger region of the $d$--$\dot{E}$ parameter space as distinguishable, allowing more halo candidates to be resolved by both CTA and LHAASO-KM2A. This behavior is expected, since the AIC is designed for relative model comparison between non-nested hypotheses by explicitly accounting for differences in model complexity, whereas the $\chi^{2}$ test primarily evaluates the absolute goodness of fit of a given model based on 1D profiles and is therefore more conservative.

Comparing the results obtained with different competing spatial templates, it is apparent that distinguishing diffusion templates from disk shapes is easier compared to differentiating them from Gaussian ones. 
This reflects the fact that disk models, characterized by flat central areas and sharp boundaries, differ more significantly from the smooth radial profiles produced by diffusion. On the other hand, a Gaussian template may partially mimic the central-peaked signature of diffusion halos, especially under conditions of restricted angular resolution or photon statistics.

The complementary performance of CTA and LHAASO-KM2A as a function of pulsar distance is clearly demonstrated. For relatively nearby objects ($d\lesssim1.5$~kpc), where the intrinsic angular extension is large, both experiments are in principle capable of resolving the morphology. In this regime, the discrimination power is mainly driven by photon statistics rather than angular resolution. Owing to its nearly continuous operation, LHAASO-KM2A accumulates substantially higher event statistics, resulting in a stronger discrimination capability than CTA for relatively nearby halos. For more distant pulsar halos, the apparent angular size becomes smaller, and the discrimination power is increasingly limited by the PSF. In this regime, CTA outperforms LHAASO-KM2A despite its shorter exposure time, owing to its superior angular resolution. Although LHAASO-KM2A benefits from larger photon statistics, its broader PSF limits its ability to resolve subtle morphological differences at large distances. Additionally, CTA's effectiveness for nearby halos is constrained by its limited field of view (FoV). For extremely extended sources like the Geminga halo, off-axis acceptance and the lack of gamma-ray-free regions for background estimation introduce significant challenges. Thus, our CTA forecasts are more reliable for compact or distant halos, while LHAASO-KM2A remains superior for nearby objects with large angular extensions due to its wide FoV advantage.

The impact of enhanced instrumental performance is also clearly demonstrated. Improving the angular resolution of LHAASO-KM2A yields only marginal gains for nearby sources, as their morphology is already well resolved and the available photon statistics remain unchanged. In contrast, at larger distances, the improved PSF substantially extends the parameter space within which halos can be discerned, enabling the detection of sources that would otherwise remain unresolved. Increasing the CTA exposure time from $50$~h to $200$~h systematically enhances discrimination capability across both nearby and distant sources, since its angular resolution is already sufficient and the additional exposure primarily boosts the photon statistics.

Consistent with current observational results, LHAASO-KM2A can presently perform unambiguous morphological discrimination only for the nearby Geminga and Monogem halos. However, with a $40\%$ improvement in angular resolution, LHAASO-KM2A would be able to identify several major pulsar halo candidates, namely the $\gamma$-ray halos surrounding pulsars J1831$‑$0952, J0248$+$6021, and J0359$+$5414. With its nominal design performance and a typical exposure of $50$~h, CTA is already capable of distinguishing the morphology of most candidate pulsar halos. The most challenging among the current candidates is the halo associated with LHAASO J0621+3755, as its pulsar has a spin-down luminosity comparable to that of Geminga but lies at a distance about six times larger. With an increased exposure of $200$~h, CTA is expected to achieve morphological discrimination between the diffusion and disk templates for LHAASO J0621$+$3755. However, distinguishing the diffusion template from a Gaussian model would require even greater photon statistics.  Furthermore, a higher pair-conversion efficiency from spin-down power leads to greater data accumulation, thereby enhancing the experimental resolving power.

Additionally, observational evidence for deviations from spherical symmetry in pulsar halos has been steadily accumulating. Specifically, an asymmetric morphology for the Geminga halo was reported in the ICRC 2023 proceedings \cite{Chen:2023icrc}, which was followed by the 2024 HAWC measurement \cite{HAWC:2024scl}, suggesting a marginal asymmetry at approximately the $2\sigma$ level. Furthermore, a significantly elongated $\gamma$-ray structure has recently been identified near the millisecond pulsar MSP J0218+4232 \cite{LHAASO:2025fxd}. On the theoretical front, such asymmetries are predicted to arise from anisotropic diffusion aligned with local magnetic fields \cite{Liu:2019zyj, Yan:2025eyr} or spatially varying diffusion coefficients induced by the local environment \cite{Fang:2025eiv}. To assess how these complex morphologies might influence our results, we performed a sensitivity test in Appendix A using mock data generated from an asymmetric template. Our findings indicate that the discrimination power actually increases when the underlying source is asymmetric, primarily due to the more pronounced structural mismatch between the asymmetric diffusion morphology and the symmetric Gaussian assumption used in the fit. Therefore, the possible asymmetry of pulsar halos may not undermine the power of the morphological discrimination, demonstrating that our method is robust.

\section{Conclusion}
\label{sec:conclu}
We have investigated the capability of CTA and LHAASO-KM2A to identify pulsar halo candidates through morphological discrimination. Using realistic mock observations derived from diffusion-driven halo templates and instrument response functions, we compared the diffusion morphology against alternative, non‑nested spatial templates, including Gaussian and disk models, employing both the AIC and the $\chi^{2}$ test for statistical decision. The AIC criterion generally provides more optimistic results than the $\chi^{2}$ test, allowing a larger fraction of halos to be distinguished from simplified morphologies. Diffusion halos are more easily separated from disk templates than from Gaussian ones, reflecting the stronger morphological contrast. 

A clear complementarity between CTA and LHAASO-KM2A is presented. For relatively nearby halos with large angular extensions, LHAASO-KM2A benefits from higher photon statistics enabled by its large effective area and continuous operation. For more distant halos with smaller apparent sizes, CTA achieves superior discrimination owing to its much better angular resolution. In addition, improving angular resolution can significantly enhance the ability to identify distant pulsar halos for LHAASO-KM2A, such as LHAASO J0249$+$6022, HESS J1831$-$098, and the candidate halo around PSR J0359$+$5414. For CTA, by adopting longer observation times, it could in principle resolve the diffusion signatures of all candidate pulsar halos within $5$~kpc, including LHAASO J0621$+$3755.

The discussion in this work is based on the ideal scenario where the object is an isolated source. If the target is contaminated by other nearby $\gamma$-ray sources, morphological discrimination could be more challenging. Furthermore, identifying diffusion signatures is a necessary but insufficient condition for confirming a pulsar halo. Other possible criteria for confirming $\gamma$-ray pulsar halos include the offset between the halo and the pulsar, the relationship between the energy required for the halo and the pulsar spin-down power, the relative sizes of the $\gamma$-ray halo and the corresponding X-ray PWN \cite{Fang:2022fof}, and a comparison between the electron energy density required to explain the halo and the typical energy density of the ISM \cite{Giacinti:2019nbu}. In practice, multi-wavelength observations are also required to determine whether a pulsar halo provides the most plausible interpretation.


\bibliography{refs}
\newpage

\appendix
\section{Asymmetric halo model}
Observational evidence for deviations from spherical symmetry in pulsar halos has been steadily accumulating. Specifically, an asymmetric morphology for the Geminga halo was reported in the ICRC 2023 proceedings \cite{Chen:2023icrc}, which was followed by the 2024 HAWC measurement \cite{HAWC:2024scl} suggesting a marginal asymmetry at approximately the $2\sigma$ level. Furthermore, a significantly elongated $\gamma$-ray structure has recently been identified near the millisecond pulsar MSP J0218+4232 \cite{LHAASO:2025fxd}. Also, some theoretical studies have proposed that such asymmetries can naturally arise from anisotropic diffusion due to the alignment of local magnetic fields \cite{Liu:2019zyj, Yan:2025eyr} or spatially varying diffusion coefficients induced by the local environment \cite{Fang:2025eiv}. To assess the impact of such asymmetry on our findings, we repeated our analysis pipeline using mock halo signals with asymmetric spatial templates to verify the robustness of our results.
The asymmetric form of halo morphology we adopted follows recent observations by HAWC \cite{HAWC:2024scl, Wu:2026cbz}. The characteristic diffusion extension $\theta_d$ varies with azimuth. Same to HAWC measurement, the halo is divided into four right-angle spherical sectors with the pulsar being the center.
In each sector, the $\gamma$-ray surface brightness is still described by the form of Eq.~(\ref{eq:theta_d}) with $\theta_d$ being different in four sectors. To ensure the consistency of the model at the center and the normalization of the morphology part, the additional factors $\lambda_i \, (i=1,2,3,4)$ are multiplied. $\lambda_i$ are constrained by the following relation.
\begin{equation}
    \begin{aligned}
        \begin{cases}
                & \lambda_i \propto 1/f_i(\theta=0) \propto \theta_{d,i}^2\\
                & \frac{1}{4} \sum_i \lambda_i = 1 \\
        \end{cases}
    \end{aligned}
\label{eq:lamda_cond}
\end{equation}
We can obtain the expression in Eq.~\ref{eq:lambda}.
\begin{equation}
    \lambda_i = \frac{\theta_{d,i}^2}{\frac{1}{4} \sum_j \theta_{d,j}^2}   
    \label{eq:lambda}
\end{equation}

\begin{figure}[!htbb]
\begin{center}
\includegraphics[width=0.6\textwidth]{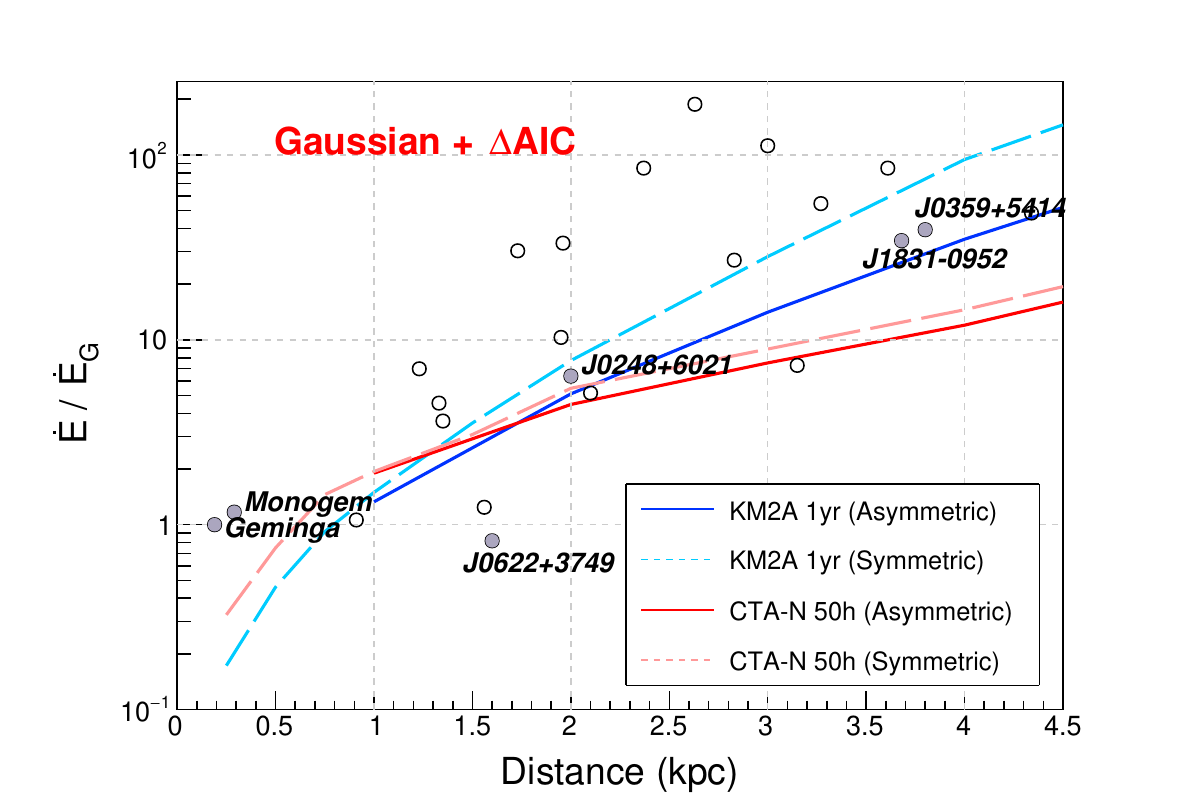}
	\caption{Morphological discrimination curves for pulsar halos based on both symmetric and asymmetric diffusion models. The curves represent the sensitivity of CTA-North and LHAASO-KM2A. The analysis employs a Gaussian spatial template as the competing model, with the AIC used as the statistical decision metric. } 
    \label{fig:asym_ts_gauss}
\end{center}
\end{figure}

For consistency, a symmetric diffusion model is still employed along with a symmetric Gaussian template to fit the mock data. $\Delta \mathrm{AIC} > 6$ is adopted as the criterion here. The new discrimination curves are presented in Fig.~\ref{fig:asym_ts_gauss}, indicating stronger discrimination capabilities in both LHAASO-KM2A and CTA,  
owing to the more pronounced deviation of the symmetric Gaussian template from the mock data, which are generated based on an asymmetric diffusion model.
The assumption of an asymmetric halo doesn't affect the main conclusions we've drawn in the main text and the complementary performance between CTA and LHAASO-KM2A remains.

\end{document}